\documentclass[twocolumn,showpacs,preprintnumbers,amsmath,amssymb,aps]{revtex4}

\usepackage{amsmath}

\usepackage{bm}
\usepackage{amssymb}
\usepackage{latexsym}

\usepackage{graphicx}

\usepackage{dcolumn}

\usepackage{bm}

\begin{document}


\title{Interacting Phantom Energy and Avoidance of the Big Rip Singularity}

\author{Ruben Curbelo, Tame Gonzalez, Genly Le\'on and Israel Quiros}

\email{rcurbelo@uclv.edu.cu;tame@uclv.edu.cu;genly@uclv.edu.cu;israel@uclv.edu.cu}

\affiliation{Universidad Central de Las Villas. Santa Clara. Cuba}

\date{\today}


\begin{abstract}
Models of the universe with arbitrary (non gravitational)
interaction between the components of the cosmic fluid: the
phantom energy and the background, are investigated. A general
form of the interaction that is inspired in scalar-tensor theories
of gravity is considered. No specific model for the phantom fluid
is assumed. We concentrate our investigation on solutions that are
free of the coincidence problem. We found a wide region in the
parameter space where the solutions are free of the big rip
singularity also. Physical arguments, together with arguments
based on the analysis of the observational evidence, suggest that
phantom models without big rip singularity might be preferred by
Nature.
\end{abstract}




\maketitle

\section{Introduction}

Recently it has been argued that astrophysical observations might
favor a dark energy (DE) component with "supernegative"
equation-of-state (EOS) parameter $\omega_{i}=p_i/\rho_i<-1$
\cite{riess,sahni}, where $p_i$ is the pressure and $\rho_i$ the
energy density of the i-th component of the cosmic fluid. Sources
sharing this property violate the null dominant energy condition
(NDEC). Otherwise, well-behaved energy sources (positive energy
density) that violate NDEC (necessarily negative pressure), have
EOS parameter less than minus unity \cite{trodden}. NDEC-violating
sources are being investigated as possible dark energy (DE)
candidates and have been called as "phantom" components
\cite{caldwell,odintsov2}.\footnotemark\footnotetext{That
NDEC-violating sources can occur has been argued decades ago, for
instance, in reference \cite{barrow}.} Since NDEC prevents
instability of the vacuum or propagation of energy outside the
light cone, then, phantom models are intrinsically unstable.
Nevertheless, if thought of as effective field theories (valid up
to a given momentum cutoff), these models could be
phenomenologically viable \cite{trodden}. Another very strange
property of phantom universes is that their entropy is
negative\cite{odintsov1}.

To the number of unwanted properties of a phantom component with
"supernegative" EOS parameter $\omega_{ph}=p_{ph}/\rho_{ph}<-1$,
we add the fact that its energy density $\rho_{ph}$ increases in
an expanding universe.\footnotemark\footnotetext{Alternatives to
phantom models to account for supernegative EOS parameter have
been considered also. See, for instance, references \cite{onemli}}
This property ultimately leads to a catastrophic (future) big rip
singularity\cite{kamionkowski} that is characterized by
divergences in the scale factor $a$, the Hubble parameter $H$ and
its time-derivative $\dot H$\cite{lazkoz}. Although other kinds of
singularity might occur in scenarios with phantom energy, in the
present paper we are interested only in the big rip kind of
singularity.\footnotemark\footnotetext{A detailed study of the
kinds of singularity might occur in phantom scenarios (including
the big rip) has been the target of reference \cite{odintsov}.}
This singularity is at a finite amount of proper time into the
future but, before it is reached, the phantom energy will rip
apart all bound structures, including molecules, atoms and nuclei.
To avoid this catastrophic event (also called "cosmic doomsday"),
some models and/or mechanisms have been invoked. In
Ref.\cite{odintsov3}, for instance, it has been shown that this
singularity in the future of the cosmic evolution might be avoided
or, at least, made milder if quantum effects are taken into
consideration. Instead, a suitable perturbation of de Sitter
equation of state can also lead to classical evolution free of the
big rip \cite{mcinnes}. Gravitational back reaction
effects\cite{wu} and scalar fields with negative kinetic energy
term with self interaction potentials with
maxima\cite{trodden,dadhich}, have also been considered .

Another way to avoid the unwanted big rip singularity is to allow
for a suitable interaction between the phantom energy and the
background (DM) fluid\cite{zhang,wang}.
\footnotemark\footnotetext{Although experimental tests in the
solar system impose severe restrictions on the possibility of
non-minimal coupling between the DE and ordinary matter fluids
\cite{will}, due to the unknown nature of the DM as part of the
background, it is possible to have additional (non gravitational)
interactions between the dark energy component and dark matter,
without conflict with the experimental data. It should be pointed
out, however, that when the stability of DE potentials in
quintessence models is considered, the coupling dark matter-dark
energy is troublesome \cite{doram}. Perhaps, the argument might be
extended to phantom models.} In effect, if there is transfer of
energy from the phantom component to the background fluid, it is
possible to arrange the free parameters of the model, in such a
way that the energy densities of both components decrease with
time, thus avoiding the big rip\cite{zhang}. Models with
interaction between the phantom and the DM components are also
appealing since the coincidence problem (why the energy densities
of dark matter and dark energy are of the same order precisely at
present?) can be solved or, at least, smoothed
out\cite{zhang,wang,pavon}.

Aim of the present paper is, precisely, to study models with
interaction between the phantom (DE) and background components of
the cosmological fluid but, unlike the phenomenological approach
followed in other cases to specify the interaction term (see
references \cite{zhang,wang}), we start with a general form of the
interaction that is inspired in scalar-tensor (ST) theories of
gravity. We concentrate our investigation on solutions that are
free of the coincidence problem. In this paper no specific model
for the phantom energy is assumed. A flat
Friedmann-Robertson-Walker (FRW) universe ($ds^2=
-dt^2+a(t)^2\delta_{ij}dx^i dx^j;\;i,j=1,2,3$) is considered that
is filled with a mixture of two interacting fluids: the background
(mainly DM) with a linear equation of state $p_m=\omega_m
\rho_m,\;\omega_m=const.$,\footnotemark\footnotetext{Although
baryons, which should be left uncoupled or weakly coupled to be
consistent with observations, are a non vanishing but small part
of the background, these are not being considered here for sake of
simplicity. Also there are suggestive arguments showing that
observational bounds on the "fifth" force do not necessarily imply
decoupling of baryons from DE\cite{pavonx}.} and the phantom fluid
with EOS parameter $\omega_{ph}=p_{ph}/\rho_{ph}<-1$. Additionally
we consider the "ratio function":
\begin{equation}
r=\frac{\rho_m}{\rho_{ph}}, \label{ratio}
\end{equation}
between the energy densities of both fluids. It is, in fact, no
more than a useful parametrization. We assume that $r$ can be
written as function of the scale factor.

The paper has been organized in the following way: In Section 2 we
develop a general formalism that is adequate to study the kind of
coupling between phantom energy and the background fluid (inspired
in ST theories as said) we want to investigate. Models of
references \cite{zhang,wang} can be treated as particular cases.
The way one can escape one of the most undesirable features of any
DE model: the coincidence problem, is discussed in Section 3 for
the case of interest, where the DE is modelled by a phantom fluid.
In this section it is also briefly discussed under which
conditions the big rip singularity might be avoided. With the help
of the formalism developed before, in Section 4, particular models
with interaction between the components of the mixture are
investigated where the coincidence problem is solved. The cases
with constant and dynamical EOS parameter are studied separately.
It is found that, in all cases, there is a wide range in parameter
space, where the solutions are free of the big rip singularity
also. It is argued that observational data seems to favor phantom
models without big rip singularity. This conclusion is reinforced
by a physical argument related to the possibility that the
interaction term is always bounded. Conclusions are given in
Section 5.

\section{General Formalism}

Since there is exchange of energy between the phantom and the
background fluids, the energy is not conserved separately for each
component. Instead, the continuity equation for each fluid shows a
source (interaction) term:
\begin{equation}
\dot\rho_m+3H(\rho_m+p_m)=Q, \label{backconteq}
\end{equation}
\begin{equation}
\dot\rho_{ph}+3H(\rho_{ph}+p_{ph})=-Q, \label{phconteq}
\end{equation}
where the dot accounts for derivative with respect to the cosmic
time and $Q$ is the interaction term. Note that the total energy
density $\rho_T=\rho_m+\rho_{ph}$ ($p_T= p_m+p_{ph}$) is indeed
conserved: $\dot\rho_T+3H(\rho_T+p_T)=0$. To specify the general
form of the interaction term, we look at a scalar-tensor theory of
gravity where the matter degrees of freedom and the scalar field
are coupled in the action through the scalar-tensor metric
$\chi(\phi)^{-1}g_{ab}$\cite{kaloper}:
\begin{eqnarray}
S_{ST}=\int d^4x
\sqrt{|g|}\{\frac{R}{2}-\frac{1}{2}(\nabla\phi)^2\nonumber\\
+\;\chi(\phi)^{-2}{\cal L}_m(\mu,\nabla\mu,\chi^{-1}g_{ab})\},
\label{staction}
\end{eqnarray}
where $\chi(\phi)^{-2}$ is the coupling function, ${\cal L}_m$ is
the matter Lagrangian and $\mu$ is the collective name for the
matter degrees of freedom. It can be shown that, in terms of the
coupling function $\chi(\phi)$, the interaction term $Q$ in
equations (\ref{backconteq}) and (\ref{phconteq}), can be written
in the following form:
\begin{equation}
Q=\rho_mH[a\frac{d(\ln{\bar\chi})}{da}], \label{interactionterm}
\end{equation}
where we have introduced the following "reduced" notation:
$\bar\chi(a)\equiv\chi(a)^{(3\omega_m-1)/2}$ and it has been
assumed that the coupling can be written as a function of the
scale factor $\bar\chi=\bar\chi(a)$.\footnotemark\footnotetext{If
$S_{ST}$ represents Brans-Dicke theory formulated in the Einstein
frame, then the coupling function $\chi(\phi)=\chi_0
\exp{(-\phi/\sqrt{\omega+3/2})}$. The dynamics of such a theory
but, for a standard scalar field with exponential self-interaction
potential has been studied, for instance, in Ref.\cite{coley}. For
a phantom scalar field it has been studied in \cite{zhang2}.} This
is the general form of the interaction we consider in the present
paper.\footnotemark\footnotetext{We recall that we do not consider
any specific model for the phantom field so, the scalar-tensor
theory given by the action $S_{ST}$ serves just as inspiration for
specifying the general form of the interaction term $Q$.}
Comparing this with other interaction terms in the bibliography,
one can obtain the functional form of the coupling function
$\bar\chi$ in each case. In Ref.\cite{zhang}, for instance,
$Q=3Hc^2(\rho_{ph}+\rho_m)= 3c^2H\rho_m(r+1)/r$, where $c^2$
denotes the transfer strength. If one compares this expression
with (\ref{interactionterm}) one obtains the following coupling
function:
\begin{equation}
\bar\chi(a)=\bar\chi_0\;e^{3\int\frac{da}{a}(\frac{r+1}{r})c^2},
\label{xizhang}
\end{equation}
where $\bar\chi_0$ is an arbitrary integration constant. If
$c^2=c_0^2=const.$ and $r=r_0=const.$, then
$\bar\chi=\bar\chi_0\;a^{3c_0^2(r_0+1)/r_0}$. Another example is
the interaction term in Ref. \cite{wang}: $Q=\delta H\rho_m$,
where $\delta$ is a dimensionless coupling function. It is related
with the coupling function $\bar\chi$ (Eqn.
(\ref{interactionterm})) through:
\begin{equation}
\bar\chi(a)=\bar\chi_0\;e^{\int\frac{da}{a} \delta},
\label{xiwang}
\end{equation}
and, for $\delta=\delta_0$;
$\bar\chi(a)=\bar\chi_0\;a^{\delta_0}$.

If one substitutes (\ref{interactionterm}) in (\ref{backconteq}),
then the last equation can be integrated:
\begin{equation}
\rho_m=\rho_{m,0}\;a^{-3(\omega_m+1)}\bar\chi,
\label{rhom}
\end{equation}
where $\rho_{m,0}$ is an arbitrary integration constant.

If one considers Eqn. (\ref{ratio}) then, Eqn.(\ref{phconteq}) can
be arranged in the form:
$\dot\rho_{ph}/\rho_{ph}+3(\omega_{ph}-1)H=(3\omega_m-1)rH[ad(\ln\chi^{-1/2})/da]$
or, after integrating:
\begin{equation}
\rho_{ph}=\rho_{ph,0}\exp\{-\int\frac{da}{a}[1+3\omega_{ph}+
ra\frac{d(\ln\bar\chi)}{da}]\},
\label{rhoph}
\end{equation}
where $\rho_{ph,0}$ is another integration constant. Using
equations (\ref{rhom}), (\ref{ratio}) and (\ref{rhoph}), it can be
obtained an equation relating the coupling function $\bar\chi$,
the phantom EOS parameter $\omega_{ph}$ and the "ratio function"
$r$:
\begin{equation}
\bar\chi(a)=\bar\chi_0(\frac{r}{r+1})
e^{-3\int\frac{da}{a}(\frac{\omega_{ph}-\omega_m}{r+1})},
\label{chi}
\end{equation}
where, as before, $\bar\chi_0$ is an integration constant.
Therefore, by the knowledge of $\omega_{ph}=\omega_{ph}(a)$ and
$r=r(a)$, one can describe the dynamics of the model under study.
Actually, if $\omega_{ph}$ and $r$ are given as functions of the
scale factor, then one can integrate in Eqn. (\ref{chi}) to obtain
$\bar\chi=\bar\chi(a)$ and, consequently, $\rho_m=\rho_m(a)$ is
given through Eqn. (\ref{rhom}). The energy density of the phantom
field can be computed through either relationship (\ref{ratio}) or
(\ref{rhoph}). Besides, the Friedmann equation can be rewritten in
terms of $r$ and one of the energy densities:
\begin{equation}
3H^2=\rho_{ph}(1+r)=\rho_m(\frac{1+r}{r}), \label{friedmanneq}
\end{equation}
so, the Hubble parameter $H=H(a)$ is also known. Due to
Eqn.(\ref{friedmanneq}), the (dimensionless) density parameters
$\Omega_i=\rho_i/3H^2$ ($\Omega_m+\Omega_{ph}=1$) are given in
terms of only $r$:
\begin{equation}
\Omega_{ph}=\frac{1}{1+r},\;\;\Omega_m=\frac{r}{1+r},
\label{densityp}
\end{equation}
respectively. It is useful to rewrite Eqn.(\ref{chi}),
alternatively, in terms of $\Omega_{ph}$ and $\omega_{ph}$:
\begin{equation}
\bar\chi(a)=\bar\chi_0\;(1-\Omega_{ph})\;
e^{-3\int\frac{da}{a}(\omega_{ph}-\omega_m)\Omega_{ph}}.
\label{chialternat}
\end{equation}
Another useful parameter (to judge whether or not the expansion if
accelerated), is the deceleration parameter $q=-(1+ \dot H/H^2)$,
that is given by the following equation:
\begin{equation}
q=-1+\frac{3}{2}[\frac{\omega_{ph}+1+(\omega_m+1)r}{1+r}],
\label{q}
\end{equation}
or, in terms of $\Omega_{ph}$ and $\omega_{ph}$:
\begin{equation}
q=\frac{1}{2}[1+3\omega_m+3(\omega_{ph}-\omega_m)\Omega_{ph}].
\label{q1}
\end{equation}

When the EOS parameter of the phantom field is a constant
$\omega_{ph}=-(1+\xi^2)<-1$, $\xi^2\in R^+$, we need to specify
only the behavior of the ratio function $r=r(a)$ or, equivalently,
of the phantom density parameter $\Omega_{ph}= \Omega_{ph}(a)$. In
this case the equation for determining of the coupling function
$\bar\chi$ (\ref{chi}) or, alternatively, (\ref {chialternat}),
can be written in either form:
\begin{equation}
\bar\chi(a)=\bar\chi_0\;(\frac{r}{r+1})\;
e^{3(1+\xi^2+\omega_m)\int\frac{da}{a(r+1)}},
\label{chiomegaconst}
\end{equation}
or,
\begin{equation}
\bar\chi(a)=\bar\chi_0\;(1-\Omega_{ph})\;
e^{3(1+\xi^2+\omega_m)\int\frac{da}{a}\Omega_{ph}}.
\label{chialternatomegaconst}
\end{equation}
The deceleration parameter is given, in this case, by $q=-1+(3/2)
[(r-\xi^2)/(r+1)]$ so, to obtain accelerated expansion
$r<2+3\xi^2$.

\section{Avoiding the coincidence problem}

In this section we discuss the way in which one of the most
undesirable features of any DE model (including the phantom) can
be avoided or evaded. For completeness we briefly discuss also the
conditions under which the doomsday event can be evaded.

\subsection{How to avoid the coincidence problem?}

It is interesting to discuss, in the general case, under which
conditions the coincidence problem might be avoided in models with
interaction among the components in the mixture. In this sense one
expects that a regime with simultaneous non zero values of the
density parameters of the interacting components is a critical
point of the corresponding dynamical system, so the system lives
in this state for a sufficiently long period of time and, hence,
the coincidence does not arises.

In the present case where we have a two-component fluid: DM plus
DE, one should look for models where the dimensionless energy
density of the background (DM) and of the DE are simultaneously
non vanishing during a (cosmologically) long period of time.
Otherwise, we are interested in ratio functions $r(a)$ that
approach a constant value $r_0\sim 1$ at late times. A precedent
of this idea but for a phenomenologically chosen interaction term
can be found, for instance, in Ref.\cite{zhang}, where a scalar
field model of phantom energy has been explored (a
self-interacting scalar field with negative kinetic energy term
$\rho_{ph}=-\dot\phi^2/2+V(\phi)$). In this case, as already
noted, the interaction term is of the form
$Q=3Hc^2(\rho_{ph}+\rho_m)= 3c^2H\rho_m(r+1)/r$. For this kind of
interaction, stability of models with constant ratio
$r_0=\rho_m/\rho_{ph}$ (and constant EOS parameter) yields to very
interesting results. In effect, it has been shown in \cite{zhang}
that, if $r_0<1$ (phantom-dominated scaling solution), the model
is stable. The solution is unstable whenever $r_0>1$
(matter-dominated scaling solution). One can trace the evolution
of the model that evolves from the unstable (matter-dominated)
scaling solution to the stable (phantom-dominated) one. As a
consequence, this kind of interaction indicates a phenomenological
solution to the coincidence problem \cite{zhang}. Actually, once
the universe reaches the stable phantom-dominated state with
constant ratio $\rho_m/\rho_{ph}= \Omega_m/\Omega_{ph}=r_0$, it
will live in this state for a very long time. Then, it is not a
coincidence to live in this long-living state where, since
$r_0\sim 1$, $\rho_m\sim\rho_{ph}$.

The above stability study can be extended to situations where no
specific model for the phantom energy is assumed (like in the case
of interest in the present paper) and where a general interaction
term $Q$ is considered (in our case $Q$ is given in
Eqn.(\ref{interactionterm})), if one introduces the following
dimensionless quantities\cite{odintsov}:
\begin{equation}
x\equiv\frac{\rho_K}{3H^2},;\;\;y\equiv\frac{\rho_V}{3H^2},
\label{dimensionless}
\end{equation}
where the "kinetic" $\rho_K\equiv (\rho_{ph}+p_{ph})/2$ and
"potential" $\rho_V\equiv (\rho_{ph}-p_{ph})/2$ terms have been
considered. Besides, if one where to consider FRW universe with
spatial curvature $k$, an additional variable:

\begin{equation}
z\equiv -\frac{k}{a^2H^2}, \label{dimensionless1}
\end{equation}
is needed. The governing equations can be written as the
three-dimensional autonomous system:
\begin{eqnarray}
x'&=&-(1+\frac{dp_{ph}}{d\rho_{ph}})(3+\frac{Q}{2H\rho_K})x\nonumber\\
&&+3x[2x+\frac{2}{3}z+(\omega_m+1)(1-x-y-z)],\nonumber\\
y'&=&-(1-\frac{dp_{ph}}{d\rho_{ph}})[3x+\frac{Q}{2H\rho_V}y]\nonumber\\
&&+3y[2x+\frac{2}{3}z+(\omega_m+1)(1-x-y-z)],\nonumber\\
z'&=&-2z+3z[2x+\frac{2}{3}z+(\omega_m+1)\nonumber\\&&(1-x-y-z)]
\label{dsystem}
\end{eqnarray}
together with the constrain $\Omega_m=1-x-y-z$, where the comma
denotes derivative with respect to the variable $N\equiv\ln a$.
The parameter range is restricted to be $0\leq x+y+z\leq 1$. In
the remaining part of this section, for simplicity, we assume that
$\omega_m=0$, i. e., the background fluid is
dust.\footnotemark\footnotetext{The details of the stability study
will be given in a separate publication, where other kinds of
interaction are considered also.}

\begin{figure}[t]
\begin{center}
\includegraphics[width=7cm]{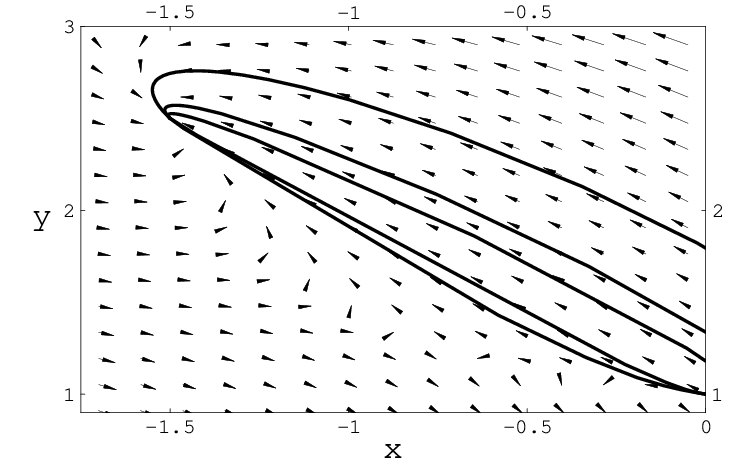}
\includegraphics[width=7cm]{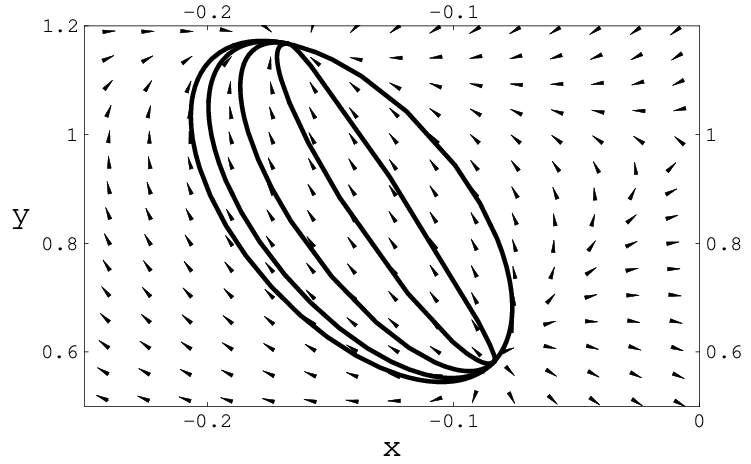}
\bigskip
\caption{The phase portraits
$(x,y)=(\frac{\rho_K}{3H^2},\frac{\rho_V}{3H^2})$ for the flat
space model with constant ratio $r_0$ and EOS parameter
$\omega_{ph,0}=-(1+\xi^2)$ are shown for background dust
($\omega_m=0$). In the upper part of the figure we have considered
$r_0<\xi^2$ ($r_0=1$, $\xi^2=3$). The point $(-3/2,5/2)$ is an
attractor. In the lower part of the figure $r_0>\xi^2$ ($r_0=1$,
$\xi^2=1/3$). All trajectories in phase space diverge from an
unstable node (third critical point $(x,y)=(-1/12,7/12)$). The
first critical point is a saddle. In any case the model has a late
time phantom dominated attractor solution (the second critical
point $(x,y)=(-1/6,7/6)$).} \label{phaseportrait}
\end{center}
\end{figure}

\subsubsection{Flat FRW ($k=0$)}

Let us consider first the spatially flat case $k=0\Rightarrow
z=0$. Consider, besides, a constant ratio $r=r_0$ and a constant
EOS parameter $\omega_{ph}=\omega_{ph,0}=-(1+\xi^2)<-1$. According
to (\ref{chiomegaconst}), one obtains a coupling function of the
form:
$\bar\chi(a)=\bar\chi_0\;(r_0/(r_0+1))\;a^{3(1+\xi^2)/(r_0+1)}$.
If one substitutes this expression for the coupling function into
Eqn. (\ref{interactionterm}), then:
\begin{equation}
Q=\frac{3(1+\xi^2)}{r_0+1}H\rho_m.
\label{interaction}
\end{equation}

The fixed points of the system (\ref{dsystem}) can be found once
the interaction function $Q$ given in Eqn.(\ref{interaction}) and
the expression $dp_{ph}/d\rho_{ph}=\omega_{ph,0}=-(1+\xi^2)$ are
substituted into (\ref{dsystem}). The fixed points of the system
$(x',y')=(0,0)$ are: 1) $(x,y)=(0,1)$, 2) $(x,y)=
(-\xi^2/2,1+\xi^2/2)$ and 3)
$(x,y)=(-\xi^2/[2(r_0+1)],(2+\xi^2)/[2(r_0+1)])$. Of these, the
first point is a saddle if $r_0>\xi^2$ and an unstable node if
$r_0<\xi^2$. The second critical point is always a stable node. In
this case $\Omega_{ph}=x+y=1$ so, there is phantom (dark energy)
dominance. The third point is an unstable node whenever
$r_0>\xi^2$ and a saddle otherwise. In this case
$\Omega_{ph}=1/(r_0+1)$ and $\Omega_m=r_0/(r_0+1)$ are
simultaneously non vanishing so, it corresponds to matter-scaling
solution. In Fig. \ref{phaseportrait} the convergence of different
initial conditions to the attractor (phantom dominated) solution
is shown in the phase space $(x,y)$ (flat space case) for given
values of the parameters $r_0$ and $\xi^2$. If $r_0>\xi^2$ all
trajectories in phase space diverge from an unstable node (third
critical point). The first critical point is a saddle. In any case
the model has a late-time phantom dominated attractor solution
(the second critical point).

\subsubsection{FRW with spatial curvature ($k\neq 0$)}

The critical points $(x'=0,y'=0,z'=0)$ are: 1) $(x,y,z)=(0,1,0)$
(de Sitter space), 2) $(x,y,z)=(-\xi^2/2,1+\xi^2/2,0)$, 3)
$(x,y,z)=(-\xi^2/[2(r_0+1)],(2+\xi^2)/[2(r_0+1)],0)$ and 4)
$(x,y,z)=(0,0,1)$ (Milne space). The first point is always a
saddle, meanwhile, the second critical point is always a stable
node (a sink). The third critical point is a saddle whenever
$r_0<\xi^2$ and an unstable node (a source) otherwise. This point
corresponds to matter-scaling solution where both $\Omega_m$ and
$\Omega_{ph}$ are simultaneously non negligible. The fourth
critical point (Milne space) corresponds to curved ($k\neq 0$)
vacuum space ($\Omega_m=\Omega_{ph}=0$). This point is an unstable
node if $r_0<2+3\xi^2$ and a saddle whenever $r_0>2+3\xi^2$. Note
that, even if the evolution of the universe begins with initial
conditions close to the fourth critical pint (curved flat space),
the model drives the subsequent evolution of the universe to the
stable node corresponding to a flat universe dominated by the
phantom component.

In summary, the way to evade the coincidence problem when an
interaction term of the general form (\ref{interactionterm}) is
considered, is then clear: One should look for models where,
whenever expansion proceeds, both $\omega_{ph}$ and the ratio
$r=\rho_m/\rho_{ph}=(1/\Omega_{ph})-1$, approach constant values
$\omega_{ph,0}$ and $r_0\sim 1$, respectively. Then, if
trajectories in phase space move close to the third critical point
(matter-scaling solution), these trajectories will stay close to
this point for an extended (arbitrarily long) period of time,
before subsequently evolving into the phantom dominated solution
(second critical point) that is a late-time attractor of the
model.\footnotemark\footnotetext{This argument is due to Alan
Coley (private communication).}

\subsection{How to avoid the big rip?}

If we assume expansion (the scale factor grows up) then, to avoid
the future big rip singularity, it is necessary that the
background fluid energy density $\rho_m$ and the phantom pressure
$p_{ph}$ and energy density $\rho_{ph}$, were bounded into the
future. Once the "ratio function" $r$ (Eqn.(\ref{ratio})) and the
parametric EOS for the phantom fluid
($p_{ph}=\omega_{ph}\rho_{ph}$) are considered, the above
requirement is translated into the requirement that $\rho_{ph}$,
$\omega_{ph}\rho_{ph}$ and $r\rho_{ph}$ were bounded into the
future. Let us consider, for illustration, the case with constant
EOS: $\omega_{ph}=\omega_{ph,0}=-(1+\xi^2)<-1$, and the following
ratio function: $r=r_0a^{-k};\;\xi^2,k\in R^+$ \cite{wang}.
Integrating in the exponent in Eqn.(\ref{chiomegaconst}) one
obtains the following expression for the coupling function:
\begin{equation}
\bar\chi(a)=\bar\chi_0\;r_0\;(a^k+r_0)^{-1+3(1+\xi^2)/k}.
\label{chiomegaconst2}
\end{equation}
The expression for the DM and phantom energy densities are then
given through equations (see equations (\ref{rhom}) and
(\ref{ratio})):
\begin{equation}
\rho_m=\bar\rho_{m,0}\;a^{-3}\;(a^k+r_0)^{-1+3(1+\xi^2)/k},
\label{rhomomegaconst2}
\end{equation}
and
\begin{equation}
\rho_{ph}=\bar\rho_{ph,0}\;a^{-3+k}\;(a^k+r_0)^{-1+3(1+\xi^2)/k},
\label{rhophomegaconst2}
\end{equation}
where $\bar\rho_{m,0}=\rho_{m,0}\bar\chi_0\;r_0$ and
$\bar\rho_{ph,0}= \rho_{m,0}\bar\chi_0$. We see that, for $a\gg
r_0^{1/k}$, $\rho_m \propto a^{-k+3\xi^2}$, while
$\rho_{ph}\propto a^{3\xi^2}$. For $\xi^2<k/3$ the DM energy
density $\rho_m$ decreases with the expansion of the universe.
However, the phantom energy density always increases with the
expansion and, consequently, the big rip singularity is
unavoidable in this example. The conclusion is that, for a phantom
fluid with a constant EOS $\omega_{ph}=\omega_{ph,0}$ and a
power-law ratio of phantom to background (dust) fluid energy
densities $\rho_m/\rho_{ph}\propto a^{-k}$, the big rip
singularity is unavoidable.

For a constant ratio $r=r_0$ and a constant EOS
($\omega_{ph}=\omega_{ph,0}=-(1+\xi^2)<-1$) \cite{zhang}, direct
integration in Eqn. (\ref{chiomegaconst}) yields to the following
coupling function:
\begin{equation}
\bar\chi(a)=\bar\chi_0\;(\frac{r_0}{r_0+1})\;a^{\frac{3(1+\xi^2)}
{r_0+1}}, \label{chiomegaconst1}
\end{equation}
so, the background and phantom energy densities in Eqn.
(\ref{rhom}) are given by:
\begin{equation}
\rho_{ph}=r_0^{-1}\rho_m=\frac{\bar\rho_{m,0}}{r_0}\;a^{-3(\frac{r_0-\xi^2}
{r_0+1})}, \label{rhomomegaconst1}
\end{equation}
where $\bar\rho_{m,0}=\rho_{m,0}\bar\chi_0\;[r_0/(r_0+1)]$. We see
that, whenever there is expansion of the universe, for
$r_0\geq\xi^2$, there is no big rip in the future of the cosmic
evolution, since $\rho_{ph}\propto
r_0\rho_{ph}\propto\omega_{ph,0}\rho_{ph}\propto
a^{-3(r_0-\xi^2)/(r_0+1)}$ are bounded into the future. In order
to get accelerated expansion, one should have $r_0<2+3\xi^2$ so,
the ratio of background (dust) energy density to phantom energy
density should be in the range $\xi^2\leq r_0<2+3\xi^2$.

\begin{figure}[t]
\begin{center}
\includegraphics[width=7cm]{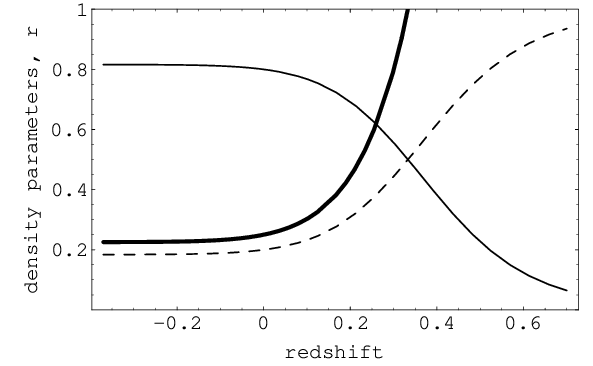}
\bigskip
\caption{The plot of energy density parameters $\Omega_m$ (dashed
curve) and $\Omega_{ph}$ (thin solid curve) as well as of the
ratio function $r$ (thick solid curve) vs redshift $z$, is shown
for the model with constant EOS parameter. For simplicity
background dust ($\omega_m=0$) was considered. The values of the
free parameters are: $m=12$, $C=0.02$ and $\omega_{ph,0}=-1.1$.
The ratio function $r$ approaches a constant value at negative
redshift $r_0=0.23$. At present ($z=0$) $r=0.25$, which means that
we are already in the long living matter-scaling state. The
background energy density dominates the early stages of the cosmic
evolution. At $z\sim 0.33$ both density parameters equate and,
since then, the phantom component dominates the dynamics of the
expansion. A very similar behavior is obtained for $\Omega_m$,
$\Omega_{ph}$ and $r$ in the case with dynamical EOS parameter
(Case B).} \label{densityp1}
\end{center}
\end{figure}

\section{Models that are free of the coincidence
problem}

In the present section we look for a model with an appropriated
coupling function $\bar\chi$ that makes possible to avoid the
coincidence problem. We study separately the cases with a constant
EOS parameter $\omega_{ph}=\omega_{ph,0}=-(1+\xi^2)<-1;\;\xi^2\in
R^+$ and with variable EOS parameter.

\subsection{Constant EOS parameter $\omega_{ph}=\omega_{ph,0}$}

In order to have models where both $\Omega_m$ and $\Omega_{ph}$
are simultaneously non-vanishing during a long interval of
cosmological time (no coincidence issue) and that, at the same
time, are consistent with observational evidence on a matter
dominated period with decelerated expansion, lasting until
recently (redshift $z\sim 0.39$ \cite{sahni}), we choose the
following dimensionless density parameter for the phantom
component:
\begin{equation}
\Omega_{ph}=\frac{m}{B}\;(\frac{a^m}{a^m+C}), \label{omegaph1}
\end{equation}
where $m$, $B$, and $C$ are arbitrary constant parameters. This
kind of density parameter function can be arranged to fit the
relevant observational data, by properly choosing the free
parameters ($m,\;B,\;C$) ($m$ controls the curvature of the curve
$\Omega_{ph}(z)$, meanwhile $C$ controls the point at which
$\Omega_{ph}(z_{eq})=\Omega_m(z_{eq})$). In this sense, the model
under study could serve as an adequate, observationally testable,
model of the universe. The choice of $\Omega_{ph}$ in equation
(\ref{omegaph1}), when substituted in Eqn.
(\ref{chialternatomegaconst}) yields to:
\begin{equation}
\bar\chi(a)=\bar\chi_0\frac{(1-\frac{m}{B})a^m+C}
{(a^m+C)^{1-3(\omega_m-\omega_{ph,0})/B}}.
\label{chi1}
\end{equation}
In consequence, equations (\ref{rhom}) and (\ref{ratio}) lead to
the following expressions determining the background energy
density and the energy density of the phantom component
respectively;
\begin{equation}
\rho_m=\bar\rho_{m,0}\;a^{-3}\;\frac{(1-\frac{m}{B})a^m+C}
{(a^m+C)^{1-3(\omega_m-\omega_{ph,0})/B}},
\label{rhom1}
\end{equation}
where $\bar\rho_{m,0}\equiv\rho_{m,0}\;\bar\chi_0/B$,
$\bar\rho_0=\bar\rho_{m,0}\bar\xi_0$ and
\begin{equation}
\rho_{ph}=\bar\rho_{ph,0}\;a^{m-3}\;(a^m+C)^{-1+3(\omega_m-\omega_{ph,0})/B},
\label{rhoph1}
\end{equation}
where $\bar\rho_{ph,0}\equiv m\;\rho_{m,0}\;\bar\chi_0/B$.

It is worth noting that, if $\omega_m-B/m\leq\omega_{ph,0}(<-1)$,
there is no big rip singularity in the future of the cosmic
evolution as described by the present model. In effect, for large
$a\gg 1$, the following magnitudes
\begin{eqnarray}
\rho_{ph}\propto r\;\rho_{ph}\propto
\omega_{ph,0}\;\rho_{ph}\nonumber\\
\propto a^{-3(1-m(\omega_m-\omega_{ph,0})/B)},
\label{asymptoticb1}
\end{eqnarray}
are bounded into the future. In particular, the phantom energy
density is a decreasing function of the scale factor. Note that
the ratio function $r=\Omega_m/\Omega_{ph}$ approaches the
constant value $(B/m)-1$ so both $\Omega_m$ and $\Omega_{ph}$ are
simultaneously non vanishing. As explained in the former section,
this means that the universe is in the third critical point
(matter-scaling solution) which is a saddle for $r_0<\xi^2$. In
consequence, the universe will evolve close to this critical point
for a sufficiently long period of time and then, the model
approaches a stable dark energy dominated regime (second critical
point).

\begin{figure}[t]
\begin{center}
\includegraphics[width=7cm]{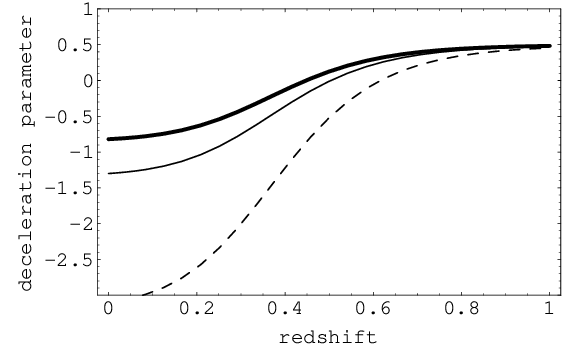}
\bigskip
\caption{The deceleration parameter $q$ is plotted as function of
the redshift $z$ for the Model A (constant EOS parameter), for
three different values of $\omega_{ph,0}=-1.1,\;-1.5$ and $-3$
(thick solid, thin solid and dashed curves respectively). Note
that the transition from decelerated (positive $q$) into
accelerated (negative $q$) expansion occurs at $z\simeq 0.4$ for
the value $\omega_{ph,0}=-1.1$. It is apparent that this is,
precisely, the observationally favored solution, since, according
to a model-independent analysis of SNIa data, $q(z=0)\approx
-0.76$\cite{john}. In consequence, solutions without big rip are
preferred by observations.} \label{decelerationp1}
\end{center}
\end{figure}

In Fig. \ref{densityp1} we show the behavior of the density
parameters $\Omega_m$ and $\Omega_{ph}$ and of the ratio function
$r$ as functions of the redshift (for simplicity we have
considered background dust: $\omega_m=0$). The following values of
the free parameters: $m=12,\;\omega_{ph,0}=-1.1,\;C=0.02$ have
been chosen, so that it is a big rip free solution. Besides we set
the DE density parameter at present epoch $\Omega_{ph}(z=0)=0.8$
so, the following relationship between the free constant
parameters $B$, $m$ and $C$ should be valid: $B=1.25[m/(1+C)]$. In
consequence, for the values of $m$ and $C$ given above: $B/m=1.22$
so, $\omega_{ph,0}=-1.1>-1.22$, meaning that there is no big rip
singularity in the future of the cosmic evolution in this model
(see the condition $-B/m\leq\omega_{ph,0}$ under Eqn.
(\ref{rhoph1})). In Fig. \ref{decelerationp1}, the plot of the
deceleration parameter $q$ vs redshift is shown for three
different values of the EOS parameter $\omega_{ph,0}=-1.1,\;-1.5$
and $-3$ respectively. It is a nice result that the solution
without big rip (thick solid curve) is preferred by the
observational evidence, since, according to a model-independent
analysis of SNIa data\cite{john}, the mean value of the present
value of the deceleration parameter is $<q_0>\approx -0.76$.

The way the coincidence problem is avoided in this model is clear
from Fig.\ref{densityp1} also. The ratio between the energy
density parameters of the background and of the DE approaches a
constant value $r_0=0.23$ for negative redshift so, both
$\Omega_m$ and $\Omega_{ph}$ are non negligible. As already said,
this is a critical point (third critical point in the stability
study of section 3) and the universe can live for a long time in
this state. Moreover, since at present ($z=0$); $r(z=0)=0.25$, we
can conclude that the universe is already in this long living
state.

\begin{figure}[t]
\begin{center}
\includegraphics[width=7cm]{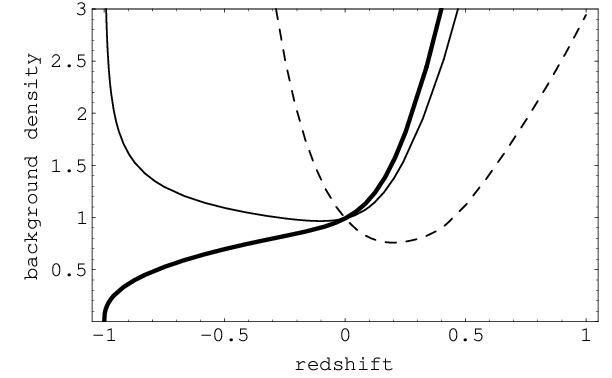}
\includegraphics[width=7cm]{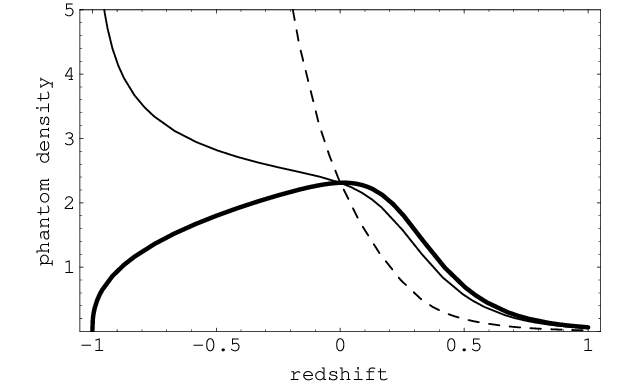}
\includegraphics[width=7cm]{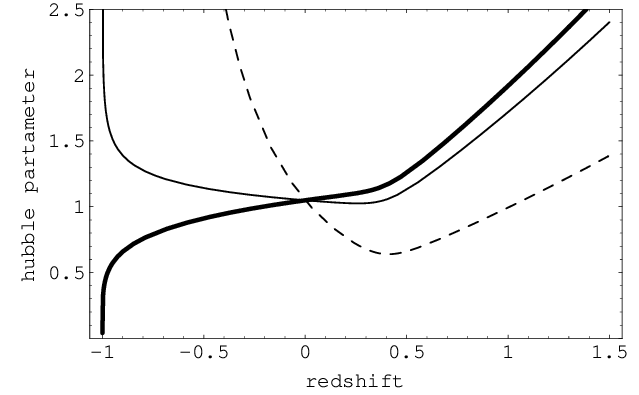}
\bigskip
\caption{The plot of the energy densities of DM $\rho_m$ (upper
part of the figure), phantom component $\rho_{ph}$ (middle part)
and of the Hubble parameter $H$ vs redshift, is shown for three
values of the constant parameter $\omega_{ph,0}$: $-1.1$ (thick
solid curve), $-1.5$ (thin solid curve) and $-3$ (dashed curve)
for the model with dynamical EOS parameter. It is apparent that,
only in the first case the model is big rip free. Note that, due
to the values chosen for the free parameters, in particular
$m=12$, the phantom energy density vanishes as one approaches the
initial big bang singularity. This is true whenever $m>3$ (see
Eqn. (\ref{rhoph2})).} \label{tresenuno}
\end{center}
\end{figure}

\begin{figure}[h]
\begin{center}
\includegraphics[width=7cm]{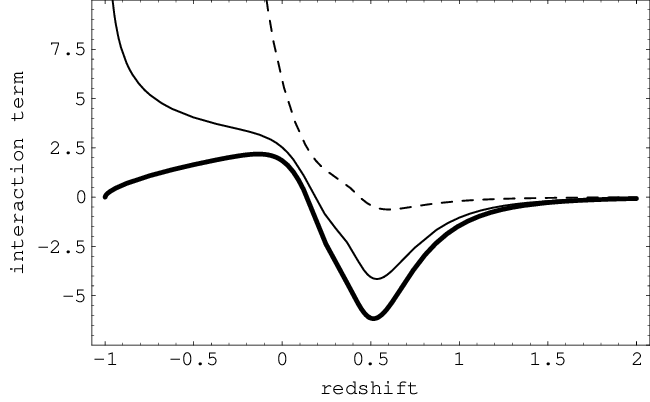}
\bigskip
\caption{The interaction term $Q$ is plotted vs redshift $z$. Note
that for $0.2\lessapprox z\lessapprox 1.7$, $Q$ is negative,
meaning that there is transfer of energy from the DM into DE
component. For higher $z$ the model proceeds without interaction.
$Q$ is bounded into the future only for the big rip free case
(thick solid curve).} \label{Q2}
\end{center}
\end{figure}

\subsection{Dynamical EOS parameter}

We now consider a dynamical EOS parameter
$\omega_{ph}=\omega_{ph}(a)$. In this case we should give as input
the functions $\Omega_{ph}(a)$ and $\omega_{ph}(a)$. In
consequence, one could integrate in the exponent in
Eqn.(\ref{chialternat}) so that the dynamics of the model is
completely specified. In order to assure avoiding of the
coincidence problem, let us consider the same phantom energy
density parameter as in the former subsection (Eqn.
(\ref{omegaph1})), but rewritten in a simpler form:
\begin{equation}
\Omega_{ph}(a)=\frac{\alpha\;a^m}{a^m+\beta}, \label{omegaph2}
\end{equation}
where $m$, $\alpha$ and $\beta$ are non negative (constant) free
parameters. To choose the function $\omega_{ph}(a)$ one should
take into account the following facts: i) at high redshift and
until recently ($z\simeq 0.39\pm 0.03$ \cite{sahni}) the expansion
was decelerated (positive deceleration parameter $q$) and, since
then, the expansion accelerates (negative $q$), ii) for negative
redshifts the EOS parameter approaches a constant value
$\omega_{ph,0}$ more negative than $-1$;
$\omega_{ph,0}=-(1+\xi^2),\;\xi^2\in R^+$. We consider,
additionally, the product $(\omega_{ph}-\omega_m)\;\Omega_{ph}$ to
be a not much complex function, so that the integral in the
exponent in Eqn.(\ref{chialternat}) could be taken analytically. A
function that fulfills all of the above mentioned requirements is
the following:
\begin{equation}
\omega_{ph}(a)=\omega_m+\omega_{ph,0}\frac{(a^m+\beta)(a^m-\delta)}{a^{2m}+\delta},
\label{omegaph2}
\end{equation}
where $\delta$ is another free parameter. The parameter $m$
controls the curvature of the density parameter function
$\Omega_{ph}(z)$, while $\delta$ controls the point of equality
$\Omega_{ph}(z_{eq})=\Omega_m(z_{eq})$. The resulting coupling
function is
\begin{eqnarray}
\bar\chi (a)=\bar\chi_0\;(1-\Omega_{ph})\;(a^{2m}+\delta)^
{-\frac{3\alpha}{2m}\omega_{ph,0}}\nonumber\\
\times\;e^{\frac{3\alpha\sqrt\delta}{m}\arctan(\frac{a^m}{\sqrt\delta})}.
\label{chi2}
\end{eqnarray}
Therefore, the background and phantom anergy densities are given
by the following expressions:
\begin{eqnarray}
\rho_m(a)=\rho_{m,0}\;[\frac{(1-\alpha)a^m+\beta}{a^m+\beta}]
\;a^{-3(\omega_m+1)}\nonumber\\
\times\;(a^{2m}+\delta)^{-\frac{3\alpha}{2m}\omega_{ph,0}}
\;e^{\frac{3\alpha\sqrt\delta}{m}\arctan(\frac{a^m}{\sqrt\delta})},
\label{rhom2}
\end{eqnarray}
\begin{eqnarray}
\rho_{ph}(a)=\rho_{ph,0}\;[\frac{a^{m-3(\omega_m+1)}}{a^m+\beta}]
\;(a^{2m}+\delta)^{-\frac{3\alpha}{2m}\omega_{ph,0}}\nonumber\\
\times\;e^{\frac{3\alpha\sqrt\delta}{m}\arctan(\frac{a^m}{\sqrt\delta})},
\label{rhoph2}
\end{eqnarray}
where the constant $\rho_{ph,0}=\alpha\;\rho_{m,0}$. At late times
(large $a$), the functions
$r\;\rho_{ph}\propto\omega_{ph}\;\rho_{ph}\propto\rho_{ph}\propto
\;a^{-3(\omega_m+1+\alpha\omega_{ph,0})}$. This means that,
whenever $-(1+\omega_m)/\alpha\leq\omega_{ph,0}(<-1)$, these
functions are bounded into the future and, consequently, there is
no big rip in the future of the cosmic evolution in the model
under study.

\begin{figure}[t]
\begin{center}
\includegraphics[width=7cm]{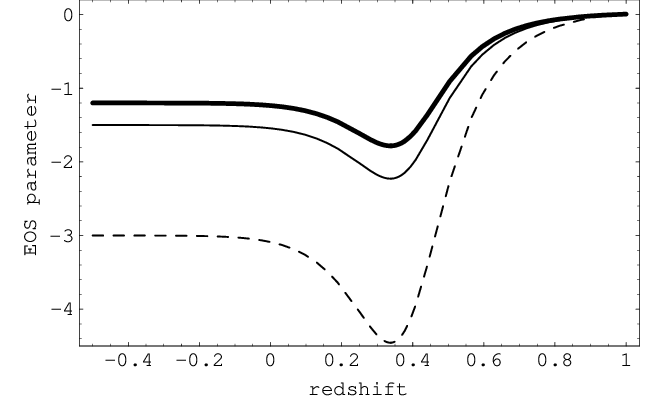}
\includegraphics[width=7cm]{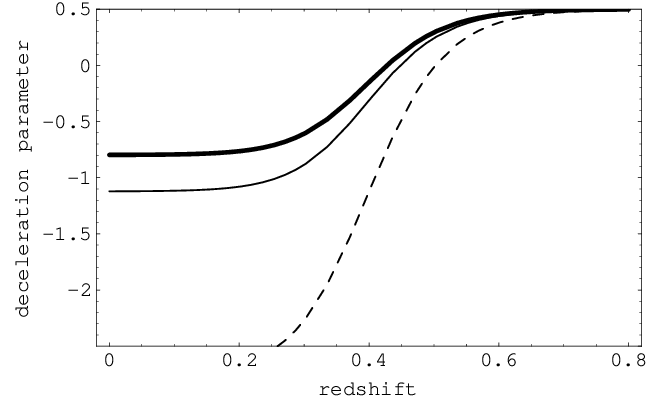}
\includegraphics[width=7cm]{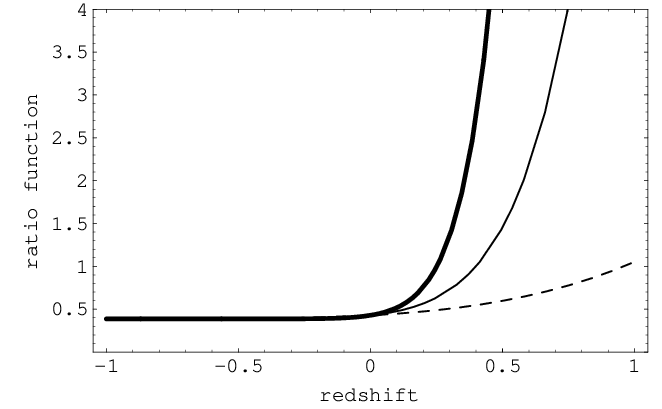}
\bigskip
\caption{In the upper and middle part of the figure the EOS
parameter $\omega_{ph}$ and the deceleration parameter $q$ are
plotted as functions of the redshift $z$ (Model B), for three
values of the constant parameter $\omega_{ph,0}=-1.1,\;-1.5$ and
$-3$ (thick solid, thin solid and dashed curves respectively). It
is worth noting that at high redshift $\omega_{ph}$ approaches a
constant value $\omega_{ph}=0$. It is a minimum at $z=0.35$. The
dark energy became a phantom fluid only recently, at $z\sim 0.45$.
Note that the transition from decelerated (positive $q$) into
accelerated (negative $q$) expansion occurs at $z\simeq 0.4$ for
the value $\omega_{ph,0}=-1.1$. This last (big rip free) solution,
seems to be preferred by observations since, according to Ref.
\cite{john}, $q(z=0)\approx -0.76$. In the lower part of the
figure the ratio function is plotted for three values of the
parameter $m$: $12$ (thick solid curve), $8$ (thin solid curve)
and $4$ (dashed curve) respectively. In any case, at present, the
universe is already in the state with constant ratio $r$.}
\label{deceEOS}
\end{center}
\end{figure}

In Fig.\ref{tresenuno} the plot of the energy densities of DM
$\rho_m$ (upper part of the figure), phantom component $\rho_{ph}$
(middle part) and of the Hubble parameter $H$ vs redshift, is
shown for three values of the constant parameter $\omega_{ph,0}$:
$-1.1$ (thick solid curve), $-1.5$ (thin solid curve) and $-3$
(dashed curve). The following values of the free parameters:
$m=12$, $\beta=0.03$, and $\delta=3\;10^{-4}$ have been chosen. We
assume $\Omega_m(z=0)=0.3$\cite{peebles} so, the following
relationship between $\alpha$ and $\beta$ should take place:
$\alpha=0.7(\beta+1)$. It is apparent that, only in the first case
($\omega_{ph,0}=-1.2$) the model is big rip free. Actually, as
already noted, the necessary requirement for absence of big rip
is, for the chosen set of parameters, $\omega_{ph,0}\geq
-1/\alpha=-1.387$. It is noticeable that, since we chose $m=12>3$,
the DE energy density diminishes as one goes back into the past
(it is a maximum at present). In all cases the interaction term is
negative during a given period of time in the past (see
Fig.\ref{Q2}), meaning that the DM transferred energy to the DE
component. At higher redshift the evolution proceeds without
interaction. Worth noting that only in the singularity free case
($\omega_{ph,0}=-1.2$) the interaction term is bounded into the
future.

A physical argument against doomsday event can be based on the
following analysis. At late time, the interaction term $Q\simeq
-3\alpha\;\omega_{ph,0}\;\rho_m\; H$. This means that, only in the
case when the evolution proceeds without big rip ($\rho_m$ and $H$
are bounded), the interaction between the components of the cosmic
fluid does not become unphysically large at finite time into the
future. Besides, for the big rip free case, as the expansion
proceeds into the future, the Hubble parameter approaches a
negligible value, meaning that the ratio $Q/\rho_m$ is small.

In the model the transition from decelerated into accelerated
expansion takes place at $z\simeq 0.4$ (see Fig. \ref{deceEOS}).
At late times the deceleration parameter approaches the constant
value $q_0\approx -0.8$. Therefore, as in the case with constant
EOS parameter (Model A), the observational evidence (as argued in
Ref. \cite{john}, $q(z=0)\approx -0.76$) seems to favor solutions
without big rip singularity. It is interesting to note that, in
this model, the dark energy component behaved like dust
($\omega_{ph}=0$) at early times (high $z$), i. e., it behaved
like an "ordinary" fluid with attractive gravity and, just until
very recently ($z\simeq 0.45$), it has not became a phantom fluid.
At present, the universe is already in a state characterized by
constant $\omega_{ph,0}\sim -1.2$ and $r_0=0.48$, meaning that
$\Omega_{m,0}=r_0/(r_0+1)$ and $\Omega_{ph,0}=1/(r_0+1)$ are
simultaneously non vanishing. As shown before, this is a critical
point of the corresponding dynamical system so, probably, the
universe will stay in this state for a long time and,
consequently, the coincidence problem does not arise.

\section{Conclusions}

In this paper we have investigated models with additional (non
gravitational) interaction between the phantom component and the
background. This kind of interaction is justified if the
interacting components are of unknown nature, as it is the case
for the DM and the DE, the dominant components in the cosmic
fluid. For the sake of simplicity, baryons, a non vanishing, but
small component of the background, have not been considered in the
present study, however, in a more complete study these should be
included also. A comment should be made in this regard: There are
suggestive arguments showing that observational bounds on the
"fifth" force do not necessarily imply decoupling of baryons from
DE\cite{pavonx}.

Unlike the phenomenological approach followed in other cases to
specify the interaction term (see references \cite{zhang,wang}),
we started with a general form of the interaction that is inspired
in ST theories of gravity (see Eqn.(\ref{interactionterm})). We
have considered no specific model for the phantom fluid. Within
this context, cosmological evolutions without coincidence problem
have been the target of this investigation. We have studied two
different models: Model A, where the DE EOS parameter is always a
constant (phantom DE) and Model B, where we considered a dynamical
DE EOS parameter (the DE became a phantom just recently). In these
models there is a wide range in the parameter space where the
cosmic evolution is free of the unwanted big rip singularity also.

The interaction between the different components of the cosmic
mixture, enables the cosmic evolution to proceed without big rip
singularity. In effect, if the interaction term is chosen so that
there is transfer of energy from the phantom component to the
background, then it is possible to arrange the free parameters of
the model to account for decreasing energy density of the phantom
fluid (as well as of the background fluid). A physical argument
against big rip can be based on the analysis of the interaction
term $Q$. In model B, for instance, at late time the interaction
term $Q\simeq -3\alpha\;\omega_{ph,0}\;\rho_m\; H$. This means
that, only whenever $\rho_m$ and $H$ are bounded (no big rip), the
interaction between the components of the cosmic fluid does not
become unphysically large at finite time into the future. Besides,
for the big rip free case, as the expansion proceeds into the
future, the Hubble parameter approaches a negligible value,
meaning that the ratio $Q/\rho_m$ is small. From the observational
perspective it seems that the evolution without doomsday event is
preferred also. For this purpose we have considered
model-independent analysis of SNIa data yielding to a mean value
of the present value of the deceleration parameter $<q_0>\approx
-0.76$\cite{john}.

Interacting models of dark energy are also useful to account for
the coincidence problem. In the present study, it has been shown
that, a solution with simultaneously non vanishing values
$\Omega_{m,0}=r_0/(r_0+1)$ and $\Omega_{ph,0}=1/(r_0+1)$ ($r_0$ is
the constant ratio of the DM to DE densities), is a critical point
of the corresponding dynamical system (third critical point in
section 3). Hence, if trajectories in phase space get close to
this point, these trajectories will stay there during a
sufficiently long period of time. This means that, once the
universe is driven into the state with simultaneously non
vanishing $\Omega_m$ and $\Omega_{ph}$, it will live in this state
for a (cosmologically) long period of time until it is finally
attracted into the phantom dominated phase being a stable node of
the models (second critical point in section 3). This is the way
the coincidence problem is solved in the present investigation. It
seems that we live already in the matter-scaling regime (third
critical point) and will live in it for long time until, finally,
the phantom dominates the destiny of the cosmic evolution. In any
case the evolution of the universe will proceed without the risk
of any catastrophic (doomsday) destiny.

An obvious limitation of the present study is that only two
specific models (model A and model B) have been considered.
However, we think this suffices to illustrate how interacting
models can be constructed that avoid the coincidence problem and ,
at the same time, fit some of the observational evidence. The
outcome that big rip free models are preferred by some of these
observations is just a nice result.

We are very grateful to Alan Coley and Sigbjorn Hervik for
valuable conversations that helped us to improve the original
version of the manuscript. We thank Marc Kamionkowski and Bret
McInnes for calling our attention upon references
\cite{kamionkowski} and \cite{mcinnes} respectively. We thank also
Joerg Jaeckel, Sergei Odintsov, Vakif Onemli, Diego Pavon and
Thanu Padmanabhan for useful comments and for pointing to us
important references. The authors thank the MES of Cuba by partial
financial support of this research.




\end{document}